\pdfoutput=1 
\documentclass[aps,prl,reprint,superscriptaddress,twocolumn,amsmath,amssymb]{revtex4-2}
\usepackage{bbding}
\usepackage[misc]{ifsym}
\usepackage{color}
\usepackage{graphicx}
\usepackage{epstopdf}
\usepackage{dcolumn}
\usepackage{bm}
\usepackage{hyperref}
\hypersetup{colorlinks=true, citecolor=blue, urlcolor=blue, linkcolor=blue}

\begin{document}

\title{Sand Creep Motion in Slow Spin-up Experiment: An Analogue of Regolith Migration on Asteroids}

\author{Chenyang Huang}
\affiliation{School of Aeronautic Science and Engineering, Beihang University, Beijing 100191, China}
\author{Yang Yu $(\textrm{\Letter})$}
\email{yuyang.thu@gmail.com}
\affiliation{School of Aeronautic Science and Engineering, Beihang University, Beijing 100191, China}

\author{Bin Cheng}
\affiliation{School of Aerospace Engineering, Tsinghua University, Beijing 100084, China}
\author{Kaiming Zhang}
\affiliation{School of Aeronautic Science and Engineering, Beihang University, Beijing 100191, China}
\author{Dong Qiao}
\affiliation{School of Aerospace Engineering, Beijing Institute of Technology, Beijing 100081, China}
\author{Hexi Baoyin}
\affiliation{School of Aerospace Engineering, Tsinghua University, Beijing 100084, China}

\begin{abstract}
We studied the creep motion of granular materials in a gradient potential field that is created using a slow spin-up experiment device. Natural sand confined in the acrylic box is spun up by a controlled turntable and the surface flows are captured using video-based measurements. Various spin-up accelerations were considered to understand the responses of creep motion on different accelerating paths. Convergent behaviors in the morphological change of sand surface were observed in the final steady state. To quantify the quasi-static spin-up process, we examined the net flux and the surface slope as a function of the spin rate and offset from the rotation axis. Evolution of sand creep motion demonstrated behaviors similar to regolith migration in numeric simulations, showing intermittency like general sheared granular systems. We noticed the sand surface approaches criticality as the spin-up proceeding, consistent with the observation that top-shaped asteroids near limiting spin rate take on critical shape. Comparisons to large-scale numeric simulations and analytical solutions reveal underlying similarities between our experiments and the million-year evolution of asteroid regolith under YORP acceleration, which raises the possibility of studying asteroid surface processes in laboratory analogue experiments.
\end{abstract}

\date{\today}

\maketitle

\begin{figure*}
    \includegraphics{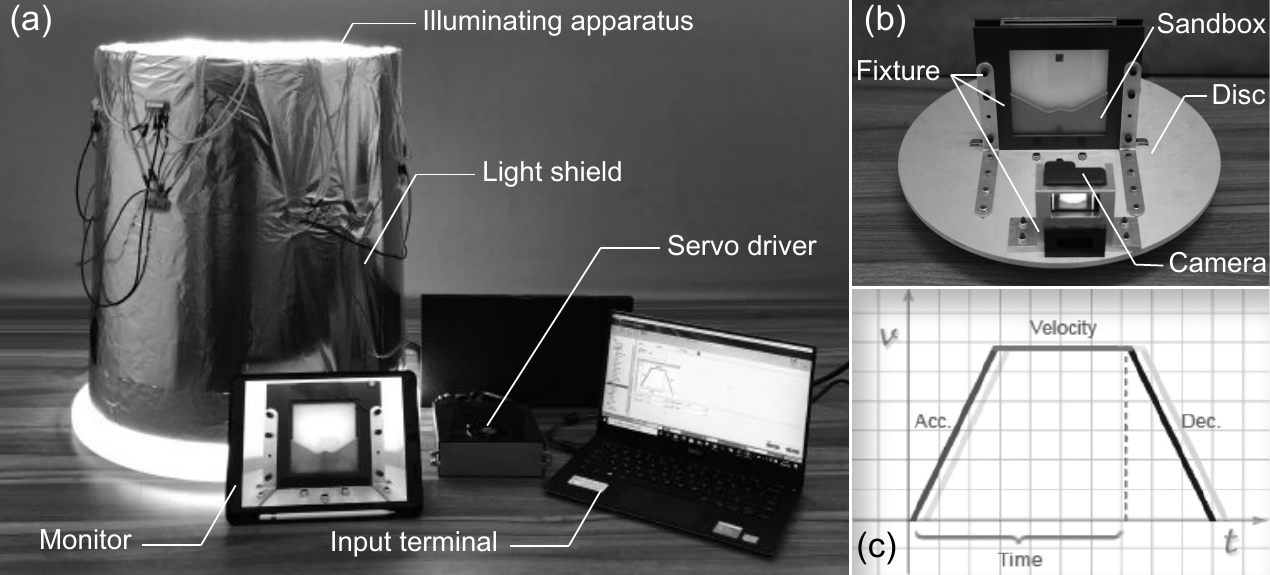}
    \centering
    \caption{\label{fig:fig1}
    The experimental setup. (a) The core components, seen in the monitor screen, are in the light shield, which prevents interference from external environment. Extra illuminating apparatus is set to provide sufficient brightness. (b) The core components. The camera and sandbox are fixed on the disc by the fixtures, keeping relatively static. (c) Control mode of the turntable and parameters to enter in the input terminal.}
\end{figure*}

\begin{figure}
  \includegraphics[width=\linewidth]{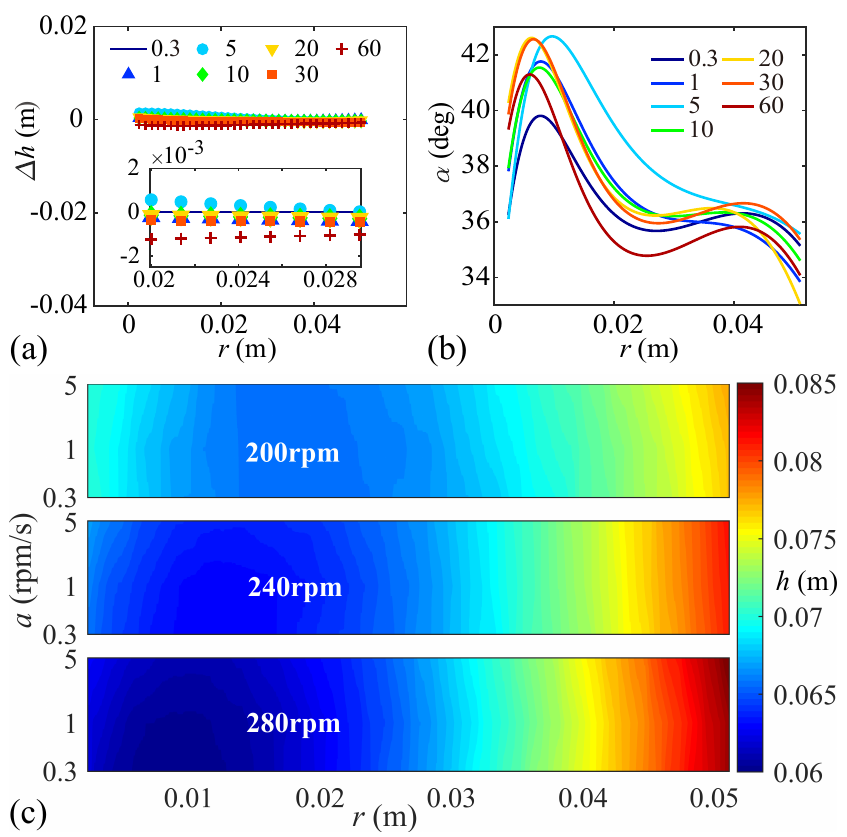}
  \centering
  \caption{\label{fig:fig2}
  (a) The relative height of sandpile in steady state as a function of the radial distance under different spin-up paths. The reference height is set as the case that the spin-up acceleration is 0.3 rpm/s. Different colors and marks indicate different accelerations. Inset: local zoom of the main graph. (b) Local slope distribution versus the radial distance. Colors indicate different accelerations. (c) Comparison of the elevation of sandpiles at the same spin rate but under different slow spin-up paths. The colorbar represents the height of sandpile. Each set of data in (a), (b) and (c) are averaged over 3 repetitions of each experiment.}
\end{figure}

\begin{figure*}
    \centering
    \includegraphics[width=\linewidth]{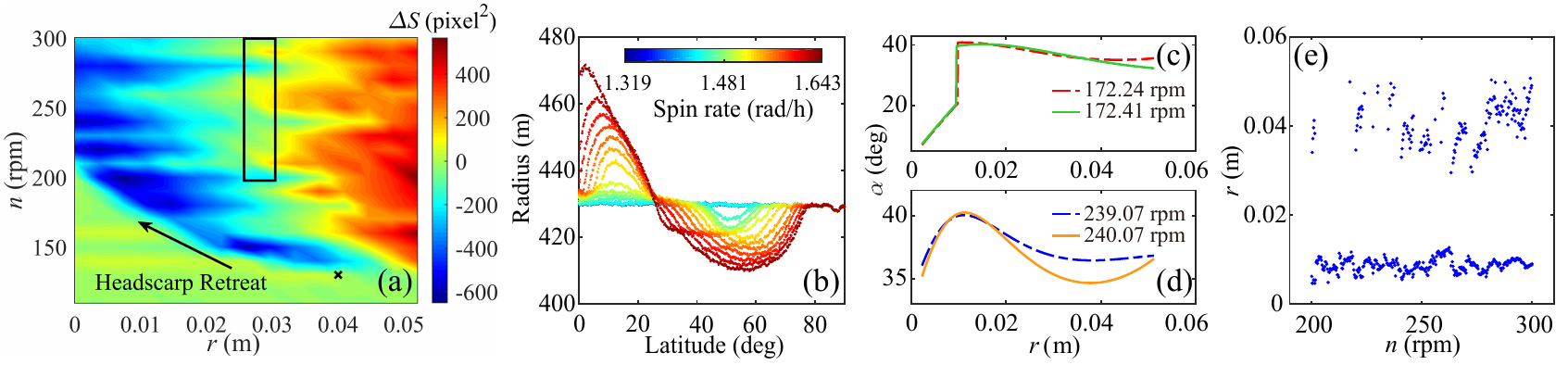}
    \caption{\label{fig:fig3}
    (a) Migration of the sand during slow spin-up process. The sandpile is divided into 15 strips in radial direction. $\rm\Delta$S of each strip is measured at set intervals (every 10 rpm during acceleration), and the space-time plot is obtained by interpolation. Color indicates $\rm\Delta$S, i.e., flow quantity of sand per unit time. The black arrow indicates the retreat of the landslide zone from the central region to the rotation axis. First creep occurs around the black marker ``$\times$''. The region in black box represents that $|\rm\Delta$S$|$ is very small (below 100 $\rm pixel^2$). (b) Variation of profile radius of top-shaped asteroid's regolith during YORP spin-up \cite{cheng2021reconstructing}. Color represents the spin rate. The profile radius is almost constant near the latitude of 26 deg and at high latitudes. (c) and (d) Typical variation of slope angle before and after landslides. (e) Locations of maximum points of local slope angle in spin-up process.}
\end{figure*}

\begin{figure}
  \centering
  \includegraphics[width=\linewidth]{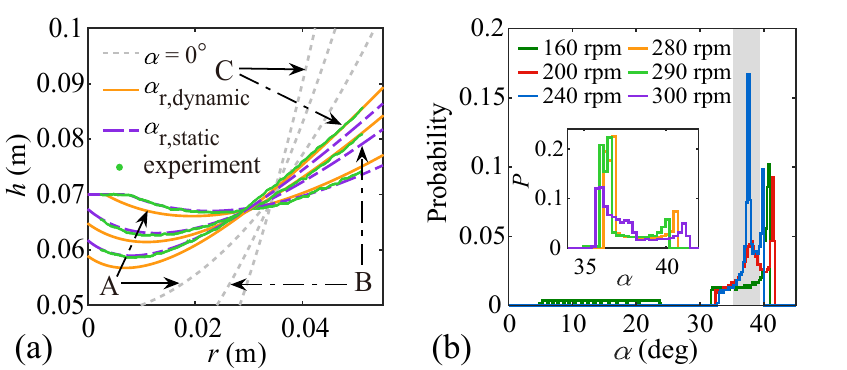}
  \caption{\label{fig:fig4}
  (a) Comparison of critical surface curves and slope surface of sandpile in experiments. Y-axis means the elevation of sandpile surface. The gray dotted line indicates the curves of slope surface using the local slope angle $\alpha$=0 everywhere, corresponding to the ideal fluid case. The orange solid line and purple dash-dot line are from the cases that we take dynamic and static angle of repose to equation (\ref{eqn2}) respectively. The green points show the slope surface of sandpile extracted from experiments. Data are from the case that angular acceleration is 0.3 rpm/s. Sets of curves, pointed by A, B and C, are at the spin rate of 180 rpm, 240 rpm and 300 rpm respectively. (b) Probability distribution of local slope angle of sandpile at different spin rates. The gray area represents the dynamic and static angle of repose limit, measured by our piling experiments. Inset: probability distribution in higher spin rates.}
\end{figure}

The creep motion of sheared granular systems has been investigated for decades using both laboratory experiments \cite{pouliquen1999scaling,komatsu2001creep,socie2005creeping,arndt2006creeping} and computational analysis \cite{silbert2001granular,richard2008rheology,dahmen2011simple,dumont2020microscopic}. But it is only in recent years that in situ space missions to Solar System small bodies have observed geological structures likely formed by the creep flows of regolith materials \cite{miyamoto2007regolith,walsh2019craters,barnouin2019shape,jawin2020global,hirabayashi2020spin}. Previous studies using DEM (Discrete Element Method) simulations have shown intriguing dynamic behaviors of the planetary scale granular systems, including overall reshaping of asteroids during YORP (Yarkovsky-O'Keefe-Radzievskii-Paddack) spin-up \cite{breiter2011yorp} and diversified surface morphologies generated by creep motions \cite{cheng2021reconstructing}. Nevertheless, the numeric methodology shows intrinsic defects in capturing the behaviours of large-scale granular systems over a long time span, which, however, are the typical characteristic of the geologic evolution on asteroids. Differing from the landslides on Earth, the regolith creep on asteroids may affect vast areas, i.e., in size comparable to the asteroid \cite{jawin2020global,morota2020sample}. The global spreading landslides lead to complex dynamics in the surface processes, especially as the asteroid is approaching the critical spin limit \cite{sanchez2018rotational,sanchez2020cohesive}. The instability in local regolith layer frequently ends with catastrophic events, like rock falling, landslides, or soil movement, which tend to redistribute the granular system into a new equilibrium \cite{barnouin2019shape,morota2020sample,scheeres2019dynamic,lauretta2019unexpected,daly2020hemispherical}. Given the long period of YORP acceleration ($\sim$ 1 Myr), such a dynamical process must be extremely slow, during which most part of the granular system maintains the critical equilibrium state, i.e., quasi-static migration. The self-organization of landslides around the criticality therefore offers a key to understand the complex behaviors of granular flows driven by dynamical disturbance on low-gravity asteroids.
\\
\indent In this letter, we designed and performed analogue experiments to mimic the creep motion of regolith material subjected to dynamical environment change on asteroids using a slowly spinning-up sandbox (Fig.\ref{fig:fig1}). By controlling the centrifugal strength, a potential gradient over the sandbox is produced to drive the creeping motion, resembling to the planetary gravity-centrifuge field in model scale. We monitor the flow dynamics in grain scale and the characteristics of sand profile shape. The results indicate that both asteroid surface evolution and our experiments are governed by similar physics, which suggests a new pathway to investigate planetary granular dynamics at a laboratory scale.
\\
\indent The experiments employ a slice acrylic sandbox (interior space 0.12$\times$0.16$\times$0.005 $\rm m^3$) filled by sand particles to a height of 0.07 m. The box is perpendicularly fixed to an aluminium alloy turning disc with sufficient stiffness to avoid deflection deformation. The field of view is 0.116 m in width as confined by the carbon fixture. Dynamic balancing is adjusted before the experiments to reduce the vibration of the platform. We choose ACW120 as the direct drive motor. The servo driver, model CDHD-0062AAF1-RO, is provided by SERVOTRONIX. This system can realize slow, smooth and enduring acceleration, with a minimum angular acceleration of 0.3 rpm/s (Revolutions Per Minute Per Second). Results are recorded with a GoPro hero 7 camera, at 60 frames per second and at a resolution of 1920$\times$1440 pixels. The camera is fixed to the support on the disc, keeping relatively static during the spin-up (Fig.\ref{fig:fig1}(b)).
\\
\indent Sand samples are collected from the Tengger Desert in the Northwest of China. Granularity detection shows a cumulative size-frequency distribution over a diameter range 40-600 $\rm\mu$m (medium size), with the best-fit power-law slope -0.7 (Supplemental Material Fig.1). The volume weighted average diameter is 184.46 $\rm\mu$m. The sandbox dimension are chosen to maintain a scale similarity to some known asteroids, i.e., the ratio of the system size to the grain diameter are of similar magnitude, which is 650:1 in our experiment, 860:1 in the regolith simulation \cite{cheng2021reconstructing}, and hundreds to thousands in some asteroids \cite{miyamoto2007regolith,walsh2019craters}. Sand piling experiments show the static angle of repose is $\sim$39.44 deg and the dynamic value is $\sim$35.23 deg (Supplemental Material Fig.2). In order to reduce wall effect of the box, we applied antistatic spray to prevent the electrostatic interference. The thickness of sandbox is set to be sufficiently large to avoid granular jamming, and meanwhile it should be small so that the sand flow maintains nearly two-dimensional. Preliminary tests show a moderate value 0.005 m is appropriate, consistent with the previous experimental values \cite{kleinhans2011static}.
\\
\indent High-quality videos of the creep-like sand flow are recorded and archived online. We placed a marker on the middle-upper part of the rectangular white mask (Supplemental Material Fig.3), then the startup time can be calibrated precisely in the video. With these settings, we are able to locate the exact video frame of a specific spin rate. We developed a geo-analyzing tool to extract the image data: 1. The full-field screen capture is denoised and then binarized, which identifies the profile of the sandpile in pixels; 2. The profile pixels are fitted using a polynomial curve fitting method (Supplemental Material Fig.4); 3. The fitted curve is analyzed frame by frame, showing the profile change of the sandpile, including the slope distribution and the flow characteristics during the creeping process.
\\
\indent Each run of the experiment includes three stages: the spin-up stage, the spin-rate-keeping stage, and the spin-down stage (Fig.\ref{fig:fig1}(c)). We focus on the first two stages. Controlled experiments were conducted for 7 parameter sets, and each experiment was repeated 3 times to reduce the influence of random factors from the environment. The system is vibrated before each experiment to make sure a initial flat packing state. We set a relatively high peek spin rate, 300 rpm, for the uniform rotation stage. Seven spin-up accelerations of the turntable are considered, i.e., 0.3, 1, 5, 10, 20, 30, and 60 rpm/s.
\\
\indent For all the spin-up accelerations as stated above, the morphological feature of the sandpile (after the turntable reaches the steady state) shows no sensitive dependence on the magnitude of acceleration. Figure \ref{fig:fig2}(a) illustrates the deviation of profile shape relative to the case that acceleration is 0.3 rpm/s. The magnitude of relative height is minuscule so the profile shapes under different spin-up paths are immensely close. The local slope angle $\alpha$ is defined as the angle between the outer normal vector of the shape curve and the effective acceleration that combines the gravitational and the centrifugal accelerations (Supplemental Material Fig.4). In all 7 experiments, the steady sandpiles exhibit an M-shape local slope change as a function of the radial distance (Fig.\ref{fig:fig2}(b)). The positions of the proximal maxima are relatively concentrated, while the distal maxima are not apparent especially under slow spin-up accelerations (\textless 10 rpm/s). Comparison of the elevation of sandpile when reaching the same spin rate through different slow spin-up paths (0.3, 1 and 5 rpm/s) is demonstrated in Fig.\ref{fig:fig2}(c). The gradient of color variation representing the elevation distribution is nearly horizontal especially at 0.3 and 1 rpm/s, which means the profile shapes of sandpile have little difference under different slow spin-up paths. Hence the two slowest spin-up paths at least could be taken as quasi-static processes.
\\
\indent Video-based technique is used to quantify the sand transport during the sandpile’s spreading in the box. The right half of the sandpile, from near rotation axis to the sidewall, is divided into 15 vertical strips. The area of sand in each strip is quantified in pixels. We define $\rm\Delta$S as the area difference of a specific strip within a selected time interval, i.e., negative $\rm\Delta$S indicates the sand is net outflow compared to the previous time point. Due to the gradient direction of the equivalent potential field (Supplemental Material Fig.5), the sand flow is unidirectional from the rotation axis to the distal side. Figure \ref{fig:fig3}(a) illustrates the spatial and temporal distribution of the net flow. It shows the first sand creep occurs near 0.04 m when the spin rate reaches $\sim$130 rpm. As the spin rate increasing, the landslide region spreads over the surface and the headscarp retreats towards the rotation axis. We have noticed a similar process of the YORP driven evolution of the regolith on asteroids described in detail by Cheng et al. \cite{cheng2021reconstructing}. In a typical YORP spin-up process, local landslide first occurs at mid-latitudes and then extends to both high and low latitudes (Supplemental Material Fig.6 and \cite{cheng2021reconstructing}). In addition, the depth of the regolith at mid-low latitudes of the asteroid shows little variation during the YORP driven process in DEM simulation (Fig.\ref{fig:fig3}(b)). As a similarity revealed by the experiments, the net flow $\rm\Delta$S of the middle strips from 0.025 to 0.03 m is close to zero for a wide speed range, which indicates an inflow-outflow balance of sand during the spin-up.
\\
\indent The fluctuations along the temporal axis correspond to a state of intermittent criticality as shown in the Supplemental Video. The large fluctuations in Fig.\ref{fig:fig3}(a) suggest that the evolution of asteroid regolith may exhibit similar statistical behaviors, although it is difficult to be captured by numerical simulations or mission cameras. The growth of the landslide region shows a swaying motion corresponding to the fluctuations, related to the self-organization of the system around the critical state: steep slope with sharp edge emerges at the moments when the sandpile is approaching critical state and subsequent landslides at scarp tend to flatten it and create a smoother landscape which is less critical (Fig.\ref{fig:fig3}(c) and Fig.\ref{fig:fig3}(d)). Based on the time-series analysis, two maximums of the slope curve are marked. The trivial maxima are located around the headscarp, suggesting a critical state of this region, while the other maxima emerge intermittently around the distal side (close to the wall) especially after the rotation rate reached 200 rpm (Fig.\ref{fig:fig3}(e)), showing the intermittency in sheared granular systems, which is also observed in previous studies \cite{lemieux2000avalanches,da2002viscosity,da2005rheophysics,arran2018intermittency}.
\\
\indent As a standard for reference, we consider an ideal profile model for the slope distribution formed by large-scale granular system, i.e., the sandbox or asteroid regolith has critical slope over the surface. Assuming the quasi-static creep motion creates a critical shape whose local slope is the angle of repose, we derived the analytical form of the profile by
\begin{eqnarray}
  <\vec{g}_e,\vec{e}_n> = \alpha_r, \label{eqn1}
\end{eqnarray}
where $\vec{g}_e$ is the effective acceleration including the gravitational and the centrifugal accelerations, $\vec{e}_n$ is the outer normal vector of slope curve, and $\alpha_r$ is the angle of repose. During the spin-up process, equation (\ref{eqn1}) controls the profile shape of the sandpile, denoted as $h=f(r)$, i.e., a function of the radial distance. Combining the boundary condition $f(r_0) = h_0$ and the mass (sectional area) conservation $\int_{r_0}^{R}f(r)dr = h_0(R-r_0)$, the analytic expression of surface elevation can be obtained:
\begin{align}
  h=&-\frac{g_0}{\omega ^2\textup{sin}^2\alpha _r}\textup{ln}\left( \omega ^2r\textup{sin}\alpha _r+g_0\textup{cos}\alpha _r \right) \nonumber\\
  &+\frac{1}{\textup{tan}\alpha _r}r+C_1. \label{eqn2}
\end{align}
In the above derivation, $R$ indicates the half of sandbox’s width, $g_0$, the gravitational acceleration on Earth, $\omega$, the angular rate, $C_1$, the integration constant, $r_0$, the radial distance where the creep motion has not occurred, and $h_0$, the corresponding hight at $r_0$ (see more details in Supplemental Material).
\\
\indent A preliminary experiment was performed to assess the constant slope angle $\alpha_r$ (Supplemental Material Fig.2), which is measured by piling sand into a cone naturally under uniform gravity. Two reference values, the dynamic and static angles of repose, can be obtained \cite{kleinhans2011static,al2018review}. Given all the parameters, we obtain the profile shapes of spinning sandpile, as shown in Fig.\ref{fig:fig4}(a) together with the experimental profile. When the sandpile is accelerated into a fast rotational speed, the experimental slope surface stays close to the critical shapes as predicted in the theoretical model using two reference values. As the spin rate increases, the local slope for most regions concentrate in a narrow range of the two angles of repose (Fig.\ref{fig:fig4}(b)). The experiments reveal that the global topography of a large-scale granular system in quasi-static spin-up meets well with the prediction of the criticality theory. This provides a benchmark result in assessing the shape evolution of fast-spinning asteroids between 100 m and 100 km, most of which are known as large-scale granular systems.
\\
\indent Likewise similarities are also noticed between the observed shapes and the theoretical shapes of small bodies limited by the angle of repose. Several small bodies near the limiting spin rates demonstrate shape features close to the critical shape, e.g., Saturnian satellite Atlas \cite{buratti2019close} shows a striking similarity to two equilibrium shapes, Donut and Top, obtained by Minton's numerical techniques \cite{minton2008topographic}. Harris et al. \cite{harris2009shapes} acquired the equilibrium solutions of a critically spinning body with constraint of constant slope at most latitudes. Choosing the angle of repose 37 deg, the ``nut-shape'' profile looks satisfyingly similar to the shape profile of 1999 KW4 Alpha \cite{scheeres2006dynamical}. Scheeres \cite{scheeres2015landslides} discussed that a rotationally symmetric body takes on uniform slopes, whose values are around angle of repose, over most of mid-latitudes and verified his prediction using observable features of asteroids such as 1999 KW4 Alpha whose slopes in mid-latitudes are in the vicinity of 40 deg \cite{ostro2006radar}. And after recomputing surface slopes of 2008 EV5 at a spin rate closer to spin limit, mid-latitudes take on a relatively uniform slope structure between 35 deg and 45 deg \cite{scheeres2015landslides,busch2011radar}. Besides, surface slopes of Ryugu and Bennu at limiting spin rates concentrate on 30 deg $\sim$ 45 deg \cite{watanabe2019hayabusa2,scheeres2016geophysical}.
\\
\indent In summary, we investigated the creep motion and landslides of a sandpile located in a gradient potential field generated by slow spin-up. Under different spin-up paths, we observed a convergent behavior in the steady-state morphology of the sandpile. We checked two extremely slow spin-up paths, 0.3 rpm/s and 1 rpm/s, which are taken as quasi-static considering the convergency in the profile of the sandpile. The surface creep occurs at mid-distal radial position first and then spreads over the whole surface, maintaining a nearly constant elevation in the middle part. The experiments show multiple similarities to the surface regolith evolution of asteroids induced by YORP spin-up in different spatial-temporal scales \cite{cheng2021reconstructing}. Intermittency of the surface sand flow is demonstrated in our experiments, suggesting similar behaviors in sheared granular systems on planetary surface. Criticality of the sandpile shape is increasingly apparent as the spin-up process proceeds, i.e., the slope angles concentrate more closely to the angle of repose. We compared our results to top-shaped asteroids approaching the spin limit, which exhibit similarities of morphological features in multiple aspects. Our results show laboratory analogue experiments of granular systems in quasi-static evolution may have great potential in studying the secular dynamical processes of planetary-scale granular systems.
\\
\begin{acknowledgments}
\indent We thank Patrick Michel for constructive conversations and useful discussions. Y. Y. acknowledges financial support provided by the National Natural Science Foundation of China Grant 12022212.
\end{acknowledgments}

% \bibliography{reference_hcy.bib}

%

\end{document}